\documentclass[fleqn,usenatbib]{mnras}

\usepackage[T1]{fontenc}

\DeclareRobustCommand{\VAN}[3]{#2}
\let\VANthebibliography\thebibliography
\def\thebibliography{\DeclareRobustCommand{\VAN}[3]{##3}\VANthebibliography}

\usepackage{graphicx}	
\usepackage{amsmath}	
\usepackage{amssymb}	
\usepackage{newtxtext,newtxmath}
\usepackage[export]{adjustbox}
\usepackage{comment}
\usepackage{soul}
\usepackage{url}

\title[Simulation of the Earth's radio-leakage from mobile towers]{Simulation of the Earth's radio-leakage from mobile towers as seen from selected nearby stellar systems}

\author[R. C. Saide et al.]{
Ramiro C. Saide,$^{1}$\thanks{E-mail: ramiro.saide@umail.uom.ac.mu}
M.A. Garrett$^{2,3}$
N. Heeralall-Issur,$^{1}$
\\
$^{1}$Department of Physics, Faculty of Science, University of Mauritius\\
$^{2}$Jodrell Bank Centre for Astrophysics, Department of Physics \& Astronomy, Alan Turing Building, The University of Manchester, M13 9PL, United Kingdom\\
$^{3}$Leiden Observatory, Leiden University, PO Box 9513, 2300 RA Leiden, The Netherlands
}

\date{Accepted XXX. Received YYY; in original form ZZZ}

\pubyear{2022}

\begin{document}
\label{firstpage}
\pagerange{\pageref{firstpage}--\pageref{lastpage}}
\maketitle

\begin{abstract}

Mobile communication towers represent a relatively new but growing contributor to the total radio leakage associated with planet Earth. We investigate the overall power contribution of mobile communication towers to the Earth's radio leakage budget, as seen from a selection of different nearby stellar systems. We created a model of this leakage using publicly available data of mobile tower locations. The model grids the surface of the planet into small, computationally manageable regions, assuming a simple integrated transmission pattern for the mobile antennas. In this model, these mobile tower regions rise and set as the Earth rotates. In this way, a dynamic power spectrum of the Earth was determined, summed over all cellular frequency bands. We calculated this dynamic power spectrum from three different viewing points - HD 95735, Barnard's star,  and Alpha Centauri A. Our preliminary results demonstrate that the peak power leaking into space from mobile towers is $\sim 4$GW. This is associated with LTE mobile tower technology emanating from the East Coast of China as viewed from HD~95735. We demonstrate that the mobile tower leakage is periodic, direction dependent, and could not currently be detected by a nearby civilisation located within 10 light years of the Earth, using instrumentation with a sensitivity similar to the Green Bank Telescope (GBT). We plan to extend our model to include more powerful 5G mobile systems, radar installations, ground based up-links (including the Deep Space Network), and various types of satellite services,  including low-Earth orbit constellations such as Starlink and OneWeb.

\end{abstract}




\section{Introduction} \label{section1}

The goal of SETI (Search for Extraterrestrial Intelligence) is to discover evidence of intelligent life beyond the Earth by looking for so-called "techno-signatures" (artificially generated signals that are not produced by nature). Unfortunately, all signals detected by SETI radio experiments to date have not been attributable to an intelligent civilisation, other than our own \citep{enriquez2017breakthrough,pinchuk2019search,traas2021breakthrough,perez2020breakthrough, wlodarczyk2020extending,price2020breakthrough,harp2016seti, heller2016search}. In principle, SETI surveys need to be sensitive to a wide range of parameter space - this is due to our ignorance regarding some very basic aspects of the signal we are looking for - including the timing of any transmissions, their location on the sky and their central frequency \citep{wright2021strategies, gray2020intermittent}. 

In parallel with the search for signs of intelligent life, the topic of exoplanets has had a major impact on the possible incidence of extraterrestrial life in the galaxy, as we understand more about the conditions on these planetary systems and their potential habitability \citep{2017AcAau.137..498W,ribas2018candidate, robinson2017characterizing}. Future advances in observing capabilities from space and on the ground have brought up new and intriguing prospects in the search for extraterrestrial life \citep{li2020opportunities, shuch2011searching, tarter2001seti}. 

Most SETI surveys are optimised to detect narrow-band signals from powerful beacons \citep{harp2011new, siemion2011developments,tarter2001search}. It is usually assumed that the detection of fainter, broad-band leakage signals is only relevant to very nearby stellar systems. The possibility of "eavesdropping" on the every-day radio transmissions inadvertently leaking into space from other technical civilisations was first considered by \citet{sullivan1978eavesdropping}. They considered the specific case of our planet Earth, and concluded that the most detectable signals were associated with military radar systems and television stations. They created a model of the Earth's leakage radiation, and demonstrated that variations in the signal power would be observed by an external observer as different regions of the Earth rotated in and out of view. 

The nature of the Earth's radio leakage has changed significantly since the pioneering work of \citet{sullivan1978eavesdropping} was published over 40 years ago. For example, powerful TV transmissions are no longer a major contributor to the Earth's leakage radiation with the rise of cable TV and the internet. In addition, mobile communication systems were unknown until the 1990s, and they currently represent a new and still growing component of the Earth's human-generated radio emission. According to \citet{stat}, the current number of mobile phone users is 7.26 billion, which suggests in excess of $\sim 91.00$\% of people in the world are cell phone owners. Individual handsets are serviced by a huge network of mobile tower systems that are spread across the landmass of the planet. Although each of these towers generates radio transmissions with relatively low power levels ($\sim$ Hundreds of watts), the directivity of these antennas and the sheer numbers involved, make them a significant component worthy of further study. It should also be noted that these mobile towers transmit at frequencies within or close to L-band, a major band for radio astronomy that includes the "water hole" defined by the natural emissions from HI and OH at $\lambda$ = 21cm and $\lambda$ = 18cm \citep{oliver1971project, 2010AcAau..67.1342S}.    

As far as we know, no previous research has investigated the cumulative effect of the mobile tower emissions and the implication for eavesdropping and SETI more generally. Other studies have demonstrated how other Earth techno-signatures would be detectable from space  at near-IR and optical wavelengths \citep{beatty2021detectability}, and from the use of power beaming to transfer energy and accelerate spacecraft \citep{benford2016power,benford2019intermediate}. Here we propose to investigate how the leakage of radiation from mobile towers on Earth would appear if viewed by an extraterrestrial civilisation using instrumentation similar to our current radio telescope technology. More specifically, we calculate the overall radio power spectrum of the Earth and the associated detection limits for a number of different exoplanet systems. We also attempt to analyse the future evolution of our mobile tower radio leakage using 5G towers as a proxy. Our study provides some insight on what we might expect if there is a human-like civilisation located elsewhere in the Milky Way with similar or indeed more advanced levels of radio telescope technology.

The paper is organised as follows. In section \ref{section2}, we provide a brief overview of the main radio communication systems currently deployed on the Earth, and details of how we model mobile communication towers in particular. We determine the integrated power spectrum of mobile towers following the methods of \citet{sullivan1978eavesdropping}. Our main results are presented in section \ref{section3}, including various power spectra for different observers located on exoplanet systems. In section \ref{section4} we explore the detectability of these emissions assuming a range of different instrumentation, including next generation telescopes such as the SKA. Our conclusions are given in section \ref{section5}, including future plans to extend our model to include additional sources of leakage radiation. 

\section{Radio leakage from the Earth.} \label{section2}

The earlier work of \citet{sullivan1978eavesdropping} identified TV transmitters and military radar systems, in particular "early-warning" radar systems, to be the most likely form of radio leakage to be detected by another intelligent civilisation observing the Earth. In particular, the detection by another civilisation of individual narrow-band TV and/or radio transmissions could be used to infer properties of our planet such as details of its rotation rate and orbital motion, information about the planet's ionosphere and troposphere, the global distribution of transmitters and possibly some cultural aspects of our civilisation.  

In this section, we consider how the Earth's radio leakage has changed in recent times, arguing that a new and important component of that emission is associated with mobile communication systems, and in particular, mobile communication towers. We have attempted to model this new component of leakage, in order to determine the radio power spectrum profile as a function of time for a specific time of the year, location of the transmitter (latitude and longitude), and celestial coordinates/distance of an outside observer (receiver).

\subsection{Earth radio mobile communication} \label{subsection1}

 The radio spectrum is used by a wide range of different services - these include: radio and television broadcasting, radio navigation and position determination, military and civilian radar systems, space and satellite communications, remote sensing applications, and so on. Over the last few decades, the development and use of mobile services by land, maritime, aeronautical, and satellite applications have grown enormously. The  frequency range used by most of these services is typically confined from 3 kHz to 30GHz, and the power levels transmitted can span many orders of magnitude ($1-10^{6}$ W) \citep{bianchi2007natural}. Each service is allocated a range of operating frequencies by the International Telecommunication Union (ITU) \citep{2000telecommunications}. 

Mobile telecommunication services have become an essential part of our modern lives, permitting us to exchange information, including video, across the planet almost instantaneously. Cellphone communication is part of  a  wider wireless communication service that has experienced exceptional growth in the last few decades. Since the early development of cell phones in 1973 \citep{murphy201340}, the number of mobile device connections has now surpassed the number of people on the planet, making it the fastest-growing human-made technology phenomenon ever. In addition, this technology is spread across the major land areas on the planet where people live. There are now over 10.98 billion mobile connections worldwide, outnumbering the current world population of 7.978 billion, according to United Nation digital analyst estimates, there are 3.002 billion more mobile connections than people on Earth \citep{stat}. 

By comparison, the number of TV stations has declined in recent years, with a massive shift towards viewing TV via cable or online streaming internet services. While TV and radio transmitters are still an important source of Earth leakage radiation at frequencies around $40-700$~MHz, there can be little doubt that mobile technologies also represent a significant part of the overall leakage budget at frequencies between $400-3000$~MHz. Currently, the most powerful sources of leakage remain military radar systems, as originally identified by \citet{sullivan1978eavesdropping}.

\subsection{Mobile Towers} \label{subsection2}

\begin{figure}
    \centering
   \includegraphics[width=1\columnwidth]{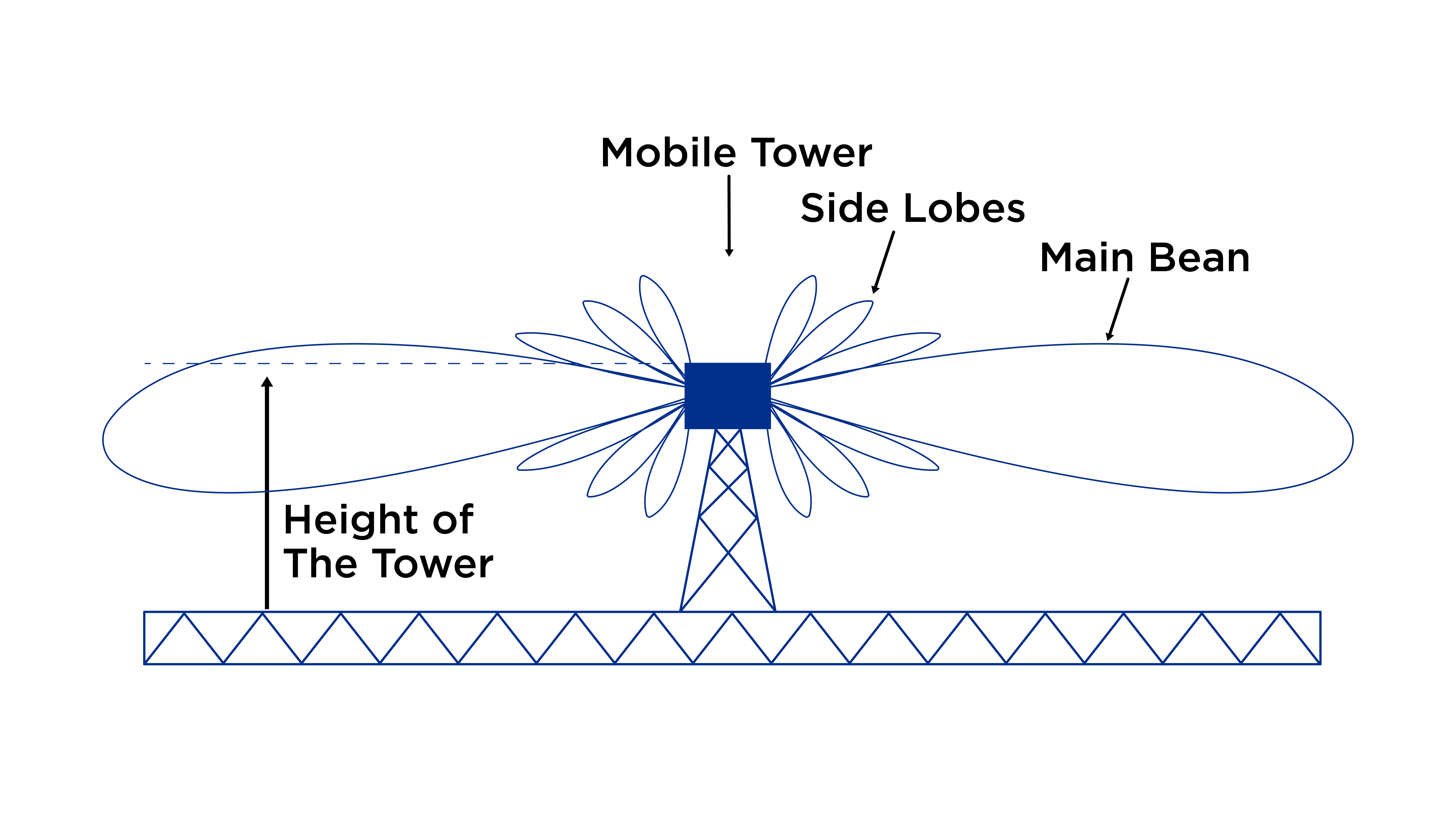}
    \caption{Radiation Pattern of a typical Mobile Tower Antenna.}
    \label{side}
\end{figure}

Mobile phones communicate by transmitting radio waves through a network of fixed antennas called mobile towers. The handsets operate at frequencies between 450 and 3000 MHz transmitting isotropically with peak powers in the range of only 0.1 to 2W  \citep{bianchi2007natural}. By comparison, a mobile tower generates peak powers of 100-200W, and the antennas are directional with significant forward gain ($\sim 50 \times$) towards the horizon (see figure \ref{side}). 

A mobile tower is a physical tower or pole for the placement of cellular radio equipment used to transmit or receive telecommunication broadcasts. The type of mobile tower is related to its purpose. Micro cell towers are smaller mobile towers that provide mobile connectivity in a small area. The distance between each micro tower is about $400-800$~m, whereas the  distance between each macro tower is typically about $2-4$~km.  The number of mobile towers in a country is dependent on the area they are required to cover, the population density and the cellular technology being deployed e.g. GSM (Global System for Mobile Communications), UMTS (Universal Mobile Telecommunications Service), LTE (Long Term Evolution), Most countries presently deploy LTE.

\subsubsection{Mobile network frequency bands} \label{subsubsection1}

The radio-frequency mobile spectrum is mainly divided into the following bands and deployed technologies:

\begin{enumerate}
    \item GSM base station antennas transmit in the frequency range of $935-960$~MHz. This frequency band of 25 MHz is divided into  twenty sub-bands  of  1.2  MHz,  which are  allocated  to various operators \citep{kumar2010cell}. There may be several carrier frequencies (1 to 5)  allotted  to  one  operator  with  an upper  limit  of  6.2  MHz bandwidth. Each carrier frequency may transmit 10 to 20W of power.  So,  one  operator  may  transmit  50  to  100W  of  power and there may be 3 - 4 operators on the same roof top or tower. Total  transmitted  power levels are therefore in the region of 200  to 400W.  In addition,  directional  antennas  are  used,  which  typically  may have  a  gain  of  around  17  dB,  so effectively,  several  kW  of  power  may  be  transmitted  in  the main beam direction \citep{kumar2010cell}. The  radiation pattern  of  directional  antennas  is  something  which  is  very critical in the whole transmission process (see figure \ref{side}).

    \item UMTS is defined as the third-generation (3G) mobile network built on the global GSM standard, and transmits in the frequency range of $1920 - 1980$~MHz, $2110 - 2170$~MHz, and these frequency bands are allocated to be used in Europe and Asia. UMTS yields channel bandwidths of 5 MHz mainly\footnote{https://www.umtsworld.com/technology/frequencies.htm} but 10 MHz and 20 MHz are also possible. UTMS presents almost the same transmitting power characteristics as GSM.

    \item  LTE yields more bands from around $600 - 3000$~MHz with a bandwidth variation from 10 MHz to greater than 100 MHz. LTE permits the following  channel bandwidths: $1.4, 3, 5, 10, 15, 20$~MHz \citep{sauter2010gsm}. The signal strength values are defined by several different measurements. \citet{6396821} showed that LTE produces more powerful radiation than UMTS and GSM. 

\end{enumerate}

\begin{figure*}
    \centering
    \includegraphics[width=1\textwidth]{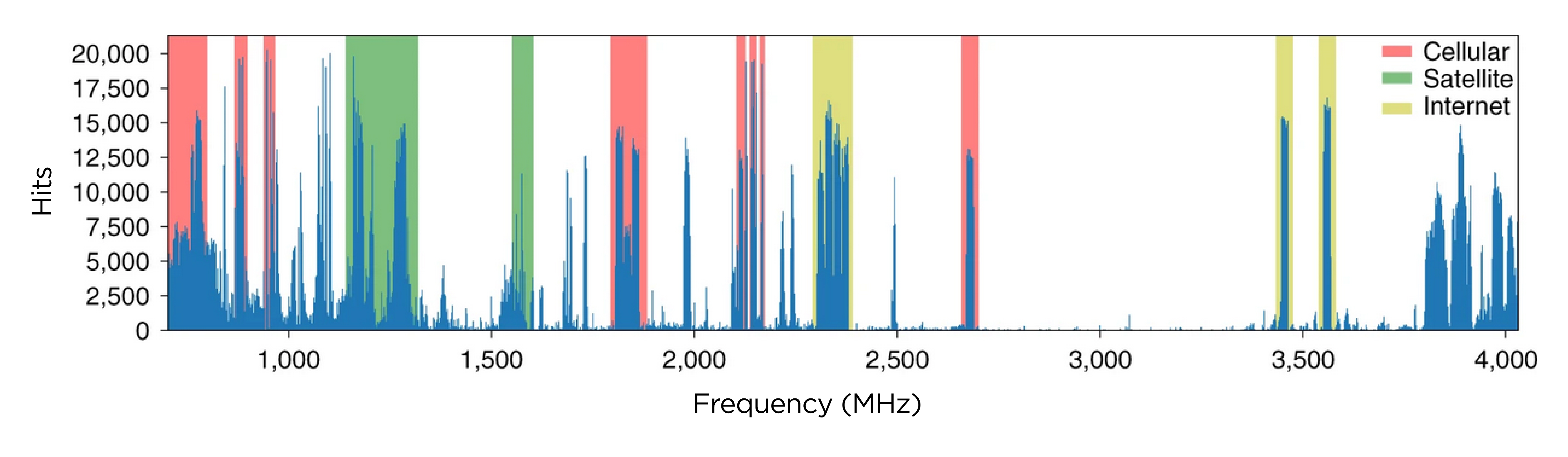}
    \caption{A histogram of total hits (signals likely to be associated with radio frequency interference) as a function of frequency \citep{smith2021radio}.}
    \label{rf}
\end{figure*}

This mobile technology can be prominently seen as a radio interference in radio observations, including recent SETI surveys e.g. \cite{smith2021radio}. In Fig.\ref{rf} of \cite{smith2021radio}, we can see the signature of the interference associated with cellular bands across the observing bandwidth. This is what an extraterrestrial observer with a very sensitive radio telescope might detect if they eavesdrop on the Earth. See also Fig.\ref{map-1} in  subsection \ref{subsection3}.

\subsection{Mobile towers Geo-location data} \label{subsection3}

We created a mobile tower database drawn from the world's largest Open Database of GPS mobile towers OpenCelliD\footnote{https://www.opencellid.org}. OpenCelliD enables public access to this database via an Application Programming Interface (API). The OpenCellID database is published under an open content license with the intention of promoting free use and redistribution of the data. The data contains different parameters, but the ones that are useful for this study include mobile tower latitude, longitude and deployed technology (GSM, UMTS and LTE). 

The OpenCellID database contains more than 30 million data points and it is updated on a daily basis. The database is populated via voluntary crowd-sourcing reports collected automatically by registered users via various mobile phone applications. There are known issues with respect to the quality of the data, in particular the data source is not complete (especially in developing countries) and may contain errors \citep{lv2019discovering, ricciato2015estimating}. Nevertheless, this is considered the best and largest, publicly available mobile tower database and it has been used in many other scientific studies \citep{johnson2021cell, ulm2015characterization, werner2019spatial}. We adopt it here, recognising that it may be incomplete in terms of geographical coverage but that it represents the best information that is currently available.

\begin{figure*}
 \centering
\includegraphics[width=1\textwidth]{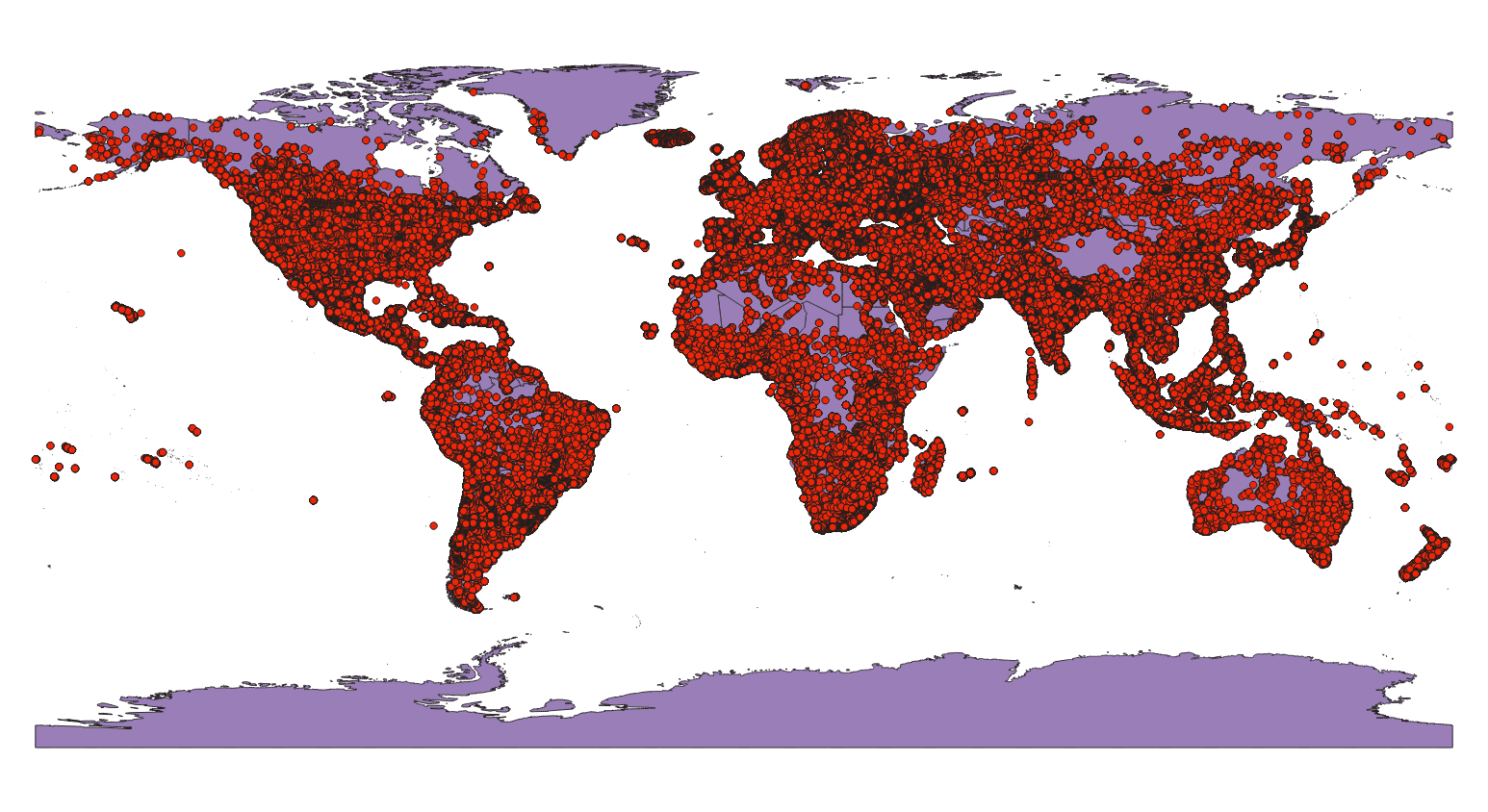}
 \caption{This map shows the geographic distribution of the mobile towers of planet Earth, represented by red dots. The map contains more than 30 million individual data points, most of which overlap at this resolution.}
 \label{map-1}
\end{figure*}

To visualize geographically the mobile towers, we used the software Qgis - Qgis is free and Open source - enabling us to create, edit, visualize, and analyze geo-spatial information. Qgis supports different types of data, such as vector, raster, delimited text, mesh layers, and many others \citep{QGIS_software, kurt2016mastering}. 

\begin{figure}
    \centering
    \includegraphics[width=\columnwidth]{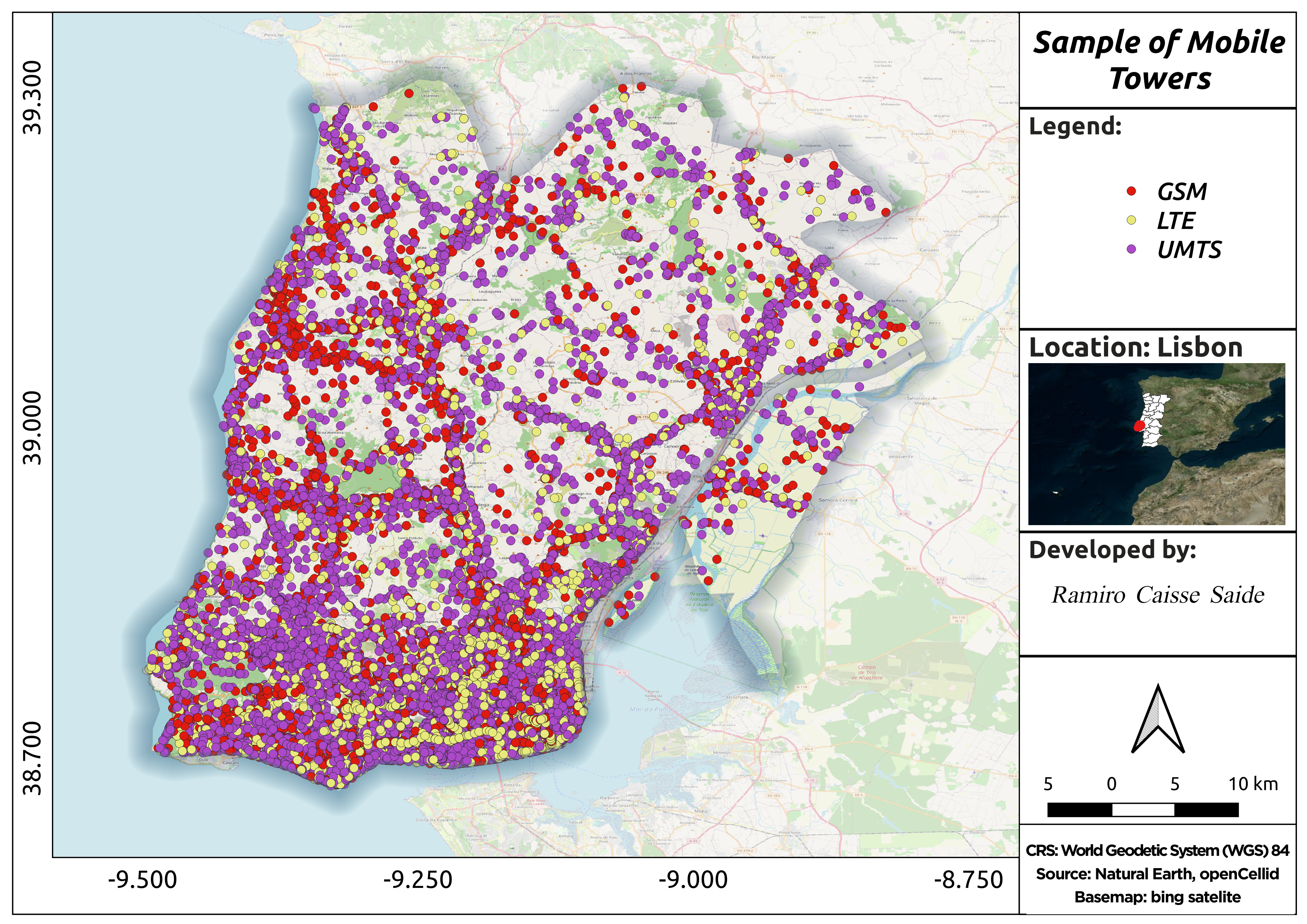}
    \caption{The distribution of mobile towers in Lisbon, Portugal. This is a small sample of the OpenCellID database available on the global scale.}
    \label{lisbon}
\end{figure}

Fig.\ref{map-1} shows the distribution of mobile towers on the Earth. The background map in purple is the Earth land coverage.
A granular view of the mobile towers can be seen in Fig.\ref{lisbon}. This sample represents the distribution of mobile towers in the city of Lisbon, Portugal. 

\begin{figure}
    \centering
    \includegraphics[width=\columnwidth]{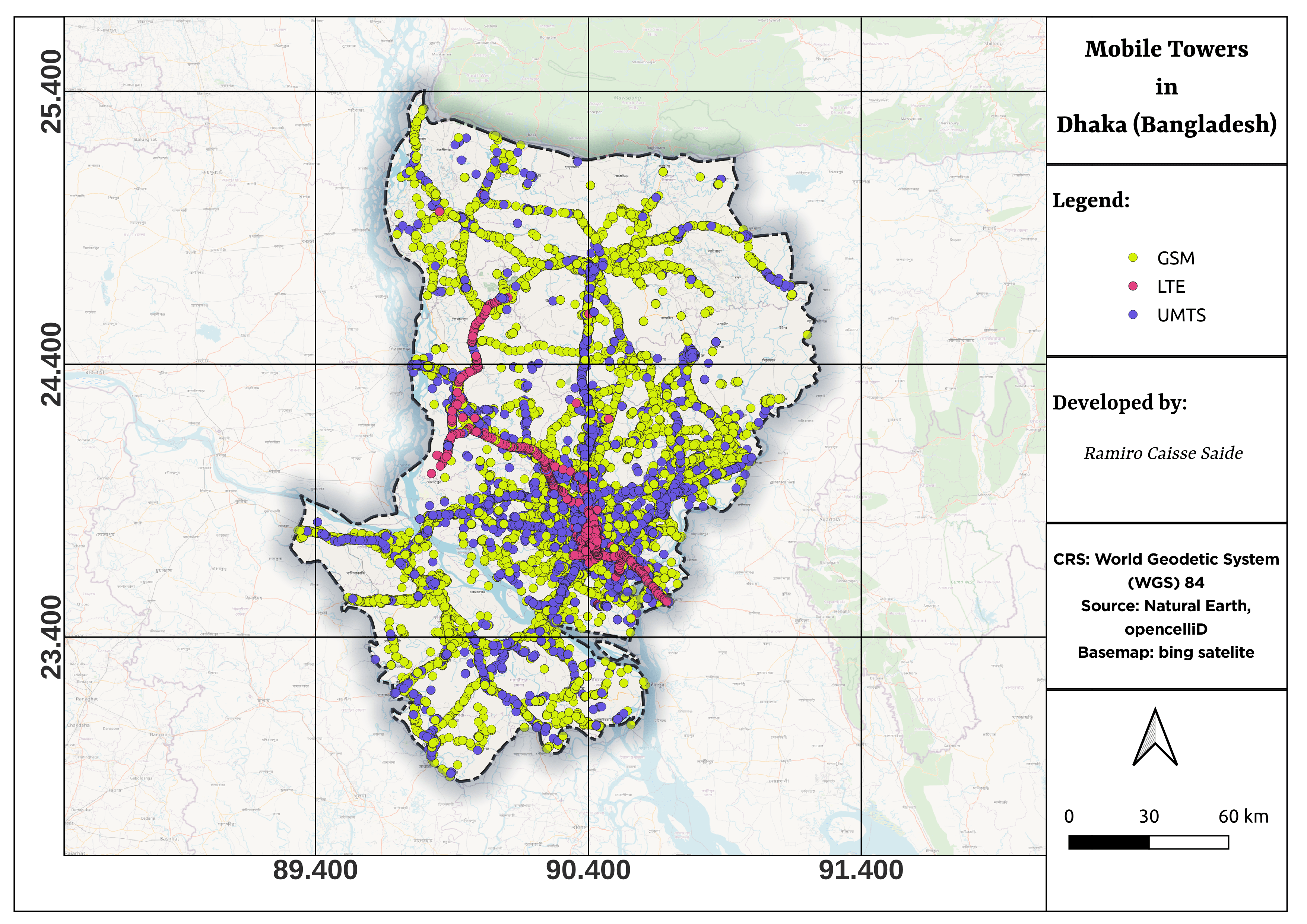}
    \caption{The distribution of the different types of mobile towers in the most densely populated city in the world, Dhaka (Bangladesh). The GSM, LTE and UMTS mobile transmitters are represented by red, yellow and purple dots respectively.}
    \label{Dhaka}
\end{figure}

A closer look at the distribution of mobile towers across the Earth, reflects the most densely populated areas and cities around the world, and provides a good sense of the granular nature of this particular form of leakage radiation. Dhaka (Bangladesh) is the world's most densely populated city (2021), with 36,941 residents per square kilometre. The distribution of mobile towers in the city can be seen in Fig.\ref{Dhaka}. There are a total of 82,833 mobile towers - this includes 45,356 GSM towers, 35,076 UMTS towers, and 2,401 LTE towers.

\subsection{Estimate of the total emitted power of mobile towers} \label{subsection4}

Mobile towers don't work with constant output powers, the power that they transmit depends on what they need to achieve. On average, a mobile tower covering a large rural area will blast out more power than a small mobile tower in the city centre. 

\begin{figure*}
    \centering
    \includegraphics[width=1\textwidth]{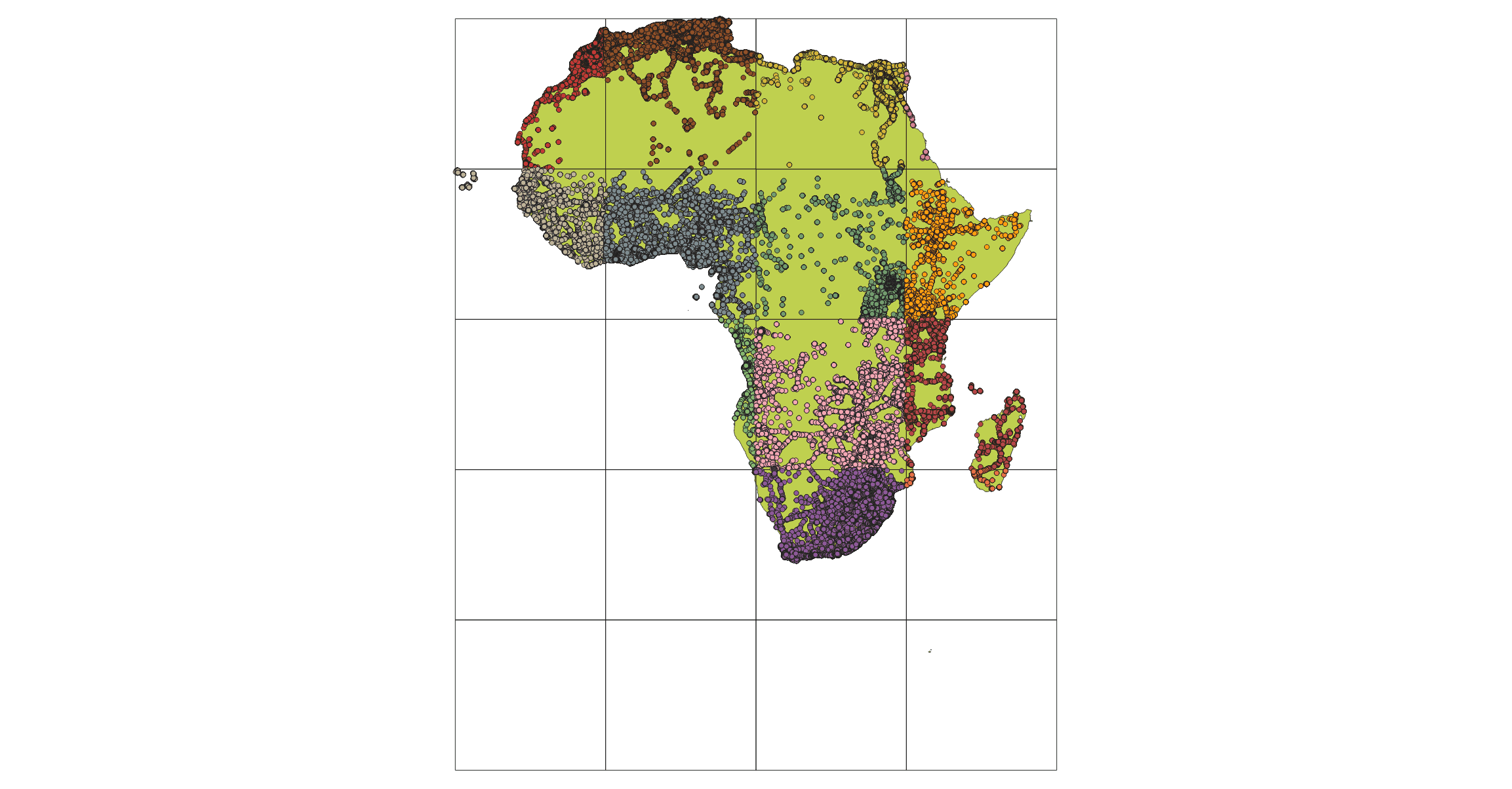}
    \caption{This image shows the towers within each grid cell for the African continent. Note that the grids are 20 degrees by 20 degrees. The centroids are calculated based on the area of the grid in the map.}
    \label{fig:data}
\end{figure*}

To estimate the total power emitted by mobile towers in great detail,  would require the emission characteristics of each individual tower to be known. Due to the amount of data we had, performing beam tracking for each individual mobile tower is computationally very expensive. In order to circumvent this problem, we divided each continent into square grids of $\sim$~20 degrees by 20 degrees (see fig.\ref{fig:data}). The grid size was effectively determined by the computing resource available to us. This approach reduced the number of mobile transmission beams that needed to be calculated but included enough granularity for the effects of the irregular geographical distribution of mobile towers to be retained in our results (see Section \ref{section3}). Fig.\ref{fig:data} shows that some grid cells contain less than 2000 towers while others contains more than 200 thousand towers.
Each central grid point became the equivalent location of the mobile tower transmitter, integrating the output of many towers into a single response. We calculated the coordinates (latitude and longitude) of the centre point of each grid - the centre of the grid encapsulated all the power of the towers within that grid. The maximum power is used as the peak of the beam's transmitter.

\begin{table*}
\centering
\begin{tabular}{l|l|l|l|l}
\hline
\textbf{Continent}       & \textbf{upper-left} & \textbf{bottom-left} & \textbf{bottom-right} & \textbf{upper-right} \\ \hline
Africa          & (-25.3604, 37.3452) & (-25.3604, -46.9657) & (51.4170, -46.9657) &    (51.4170, 37.3452) \\
\hline 
NA &    (-179.1435, 83.6341)        &   (-179.1435, -0.3887)          &   (179.7809, -0.3887)           &    (179.7809, 83.6341)         \\ \hline
SA &    (-109.4537, 15.7029)        &   (-109.4537, -55.9185)          & (-28.8770, -55.9185)  &     (-28.8770, 15.7029)        \\ \hline
Asia            &    (25.6632, 55.4345)     &  (25.6632, -12.1999)           &  (153.9856,-12.1999)            &     (153.9856,55.4345)        \\ \hline
Europe          &   (-27.3746, 84.5116)         &      (-27.3746, 16.3208)       &    (179.9843, 16.3208)   &   (179.9843, 84.5116)        \\ \hline
Oceania         &   (-180.0, 20.5554)         &   (-180.0, -54.7504)          &  (180.0, -54.7504)            &   (180.0, 20.5554)          \\ \hline
\end{tabular}
\caption{Polygon coordinates to create grids for the continents, each of these are pairs of (longitude, latitude). NA and SA represents North America and South America respectively.}
\label{table1}
\end{table*}

We calculated the contribution of each mobile tower technology deployed within each grid cell. The total power was then calculated based on the single tower Equivalent Radiated Power (ERP). We summed all the towers within each grid and multiplied by the power emitted by a single cell tower. Cellular facilities, by contrast, use a few hundred watts of EIRP per channel, depending on the purpose at any given time and the number of service providers co-located at any given tower \citep{levitt2010biological}. For simplicity, we assumed for each type of mobile technology the following output power values: GSM 100 Watts, UMTS 100 Watts, and LTE 200 Watts.

A simplified model of a beam pattern of a mobile tower was adopted using a Gaussian function - we assigned all the beam patterns as omnidirectional in azimuth and of Gaussian shape in the elevation angle above the horizon. Note that our analysis does not take into consideration reflections from buildings, mountains or other structures that reflect radio waves.

\begin{figure*}
\centering
    \includegraphics[width=1\textwidth]{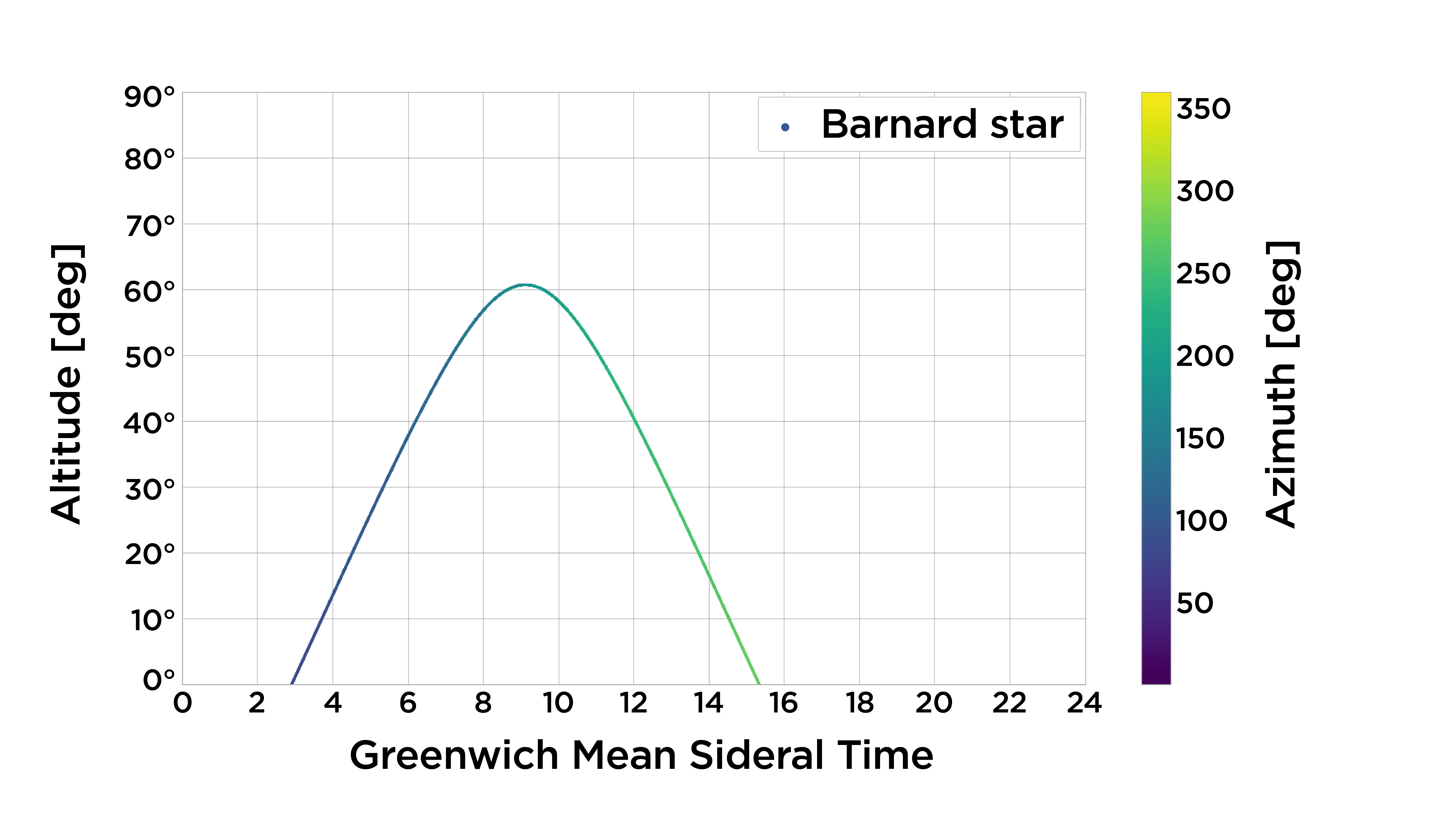}
    \caption{The position of Barnard star over 24 hours GMST from one of the grid centres located in Japan (longitude = 132.786, latitude = 34.033).}
    \label{AltAzjapan}
\end{figure*}

For each of the grid centroids, we calculated the elevation and azimuth of an extraterrestrial observer (see section \ref{section3}) for a particular day of the year (see for example, Fig.\ref{AltAzjapan}). We used the elevation angle to calculate the gain of the Gaussian beam pattern with the amplitude being the sum of the power from the towers within the respective grid cell. Finally, we calculated the total power and plotted this against Greenwich Mean Sidereal Time (GMST).

\begin{figure}
    \includegraphics[width=\columnwidth]{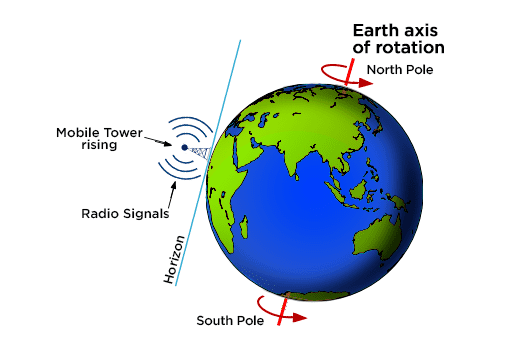}
    
    \caption{An extraterrestrial observer will detect the maximum radiation from a mobile tower as it rises or sets. This illustration shows a tower rising across the horizon.}
\label{illustration1}
\end{figure}

Mobile towers emit the maximum power towards the horizon, therefore the extraterrestrial observer  will detect the maximum power when the mobile towers are rising or setting, in other words, when the observer is on the local horizon of the mobile tower (see Fig.\ref{AltAzjapan} and Fig.\ref{illustration1}). Nevertheless, other locations also contribute to the overall detectable power - our calculations also take this aspect into account. 

\section{Results} \label{section3}

Our sample of stars was chosen primarily to be located nearby and also in terms of which coordinates of the stars could receive maximum detectable leakage. We choose one star with a declination near the equator (Barnard star), one in the southern hemisphere (HD 95735), and another in the northern hemisphere (Alpha Centauri A). Northern stars will detect more leakage radiation than southern stars due to the amount of transmitters in the northern hemisphere illuminating these stars. Last but not least, we considered also systems with potentially habitable planets, which is the case of Barnard star \citep{ribas2018candidate}.

Following \citet{sullivan1978eavesdropping}, we first determined the radio power spectrum from the Earth using the Barnard star as an extraterrestrial location. This star is a red dwarf at a distance of $\sim 6$ light-years from Earth in the constellation of Ophiuchus with a Right Ascension: $17^{h} 57^{m} 48.49803^{s}$ and Declination: $04^\circ 41' 36.2072"$ (J2000). It is the fourth-nearest-known individual star to the Sun. 

We determined the total power separately for LTE, GSM, and UMTS mobile technologies. Our preliminary results demonstrate that for Barnard star, the peak power from mobile towers with LTE technology is in the order of $\sim 3.2$~GW, out-powering the other 2 mobile technologies with peak powers of 1.3~GW for GSM and 2.7~GW for UMTS mobile towers. 

\begin{figure*}
\centering
\includegraphics[width=1\textwidth]{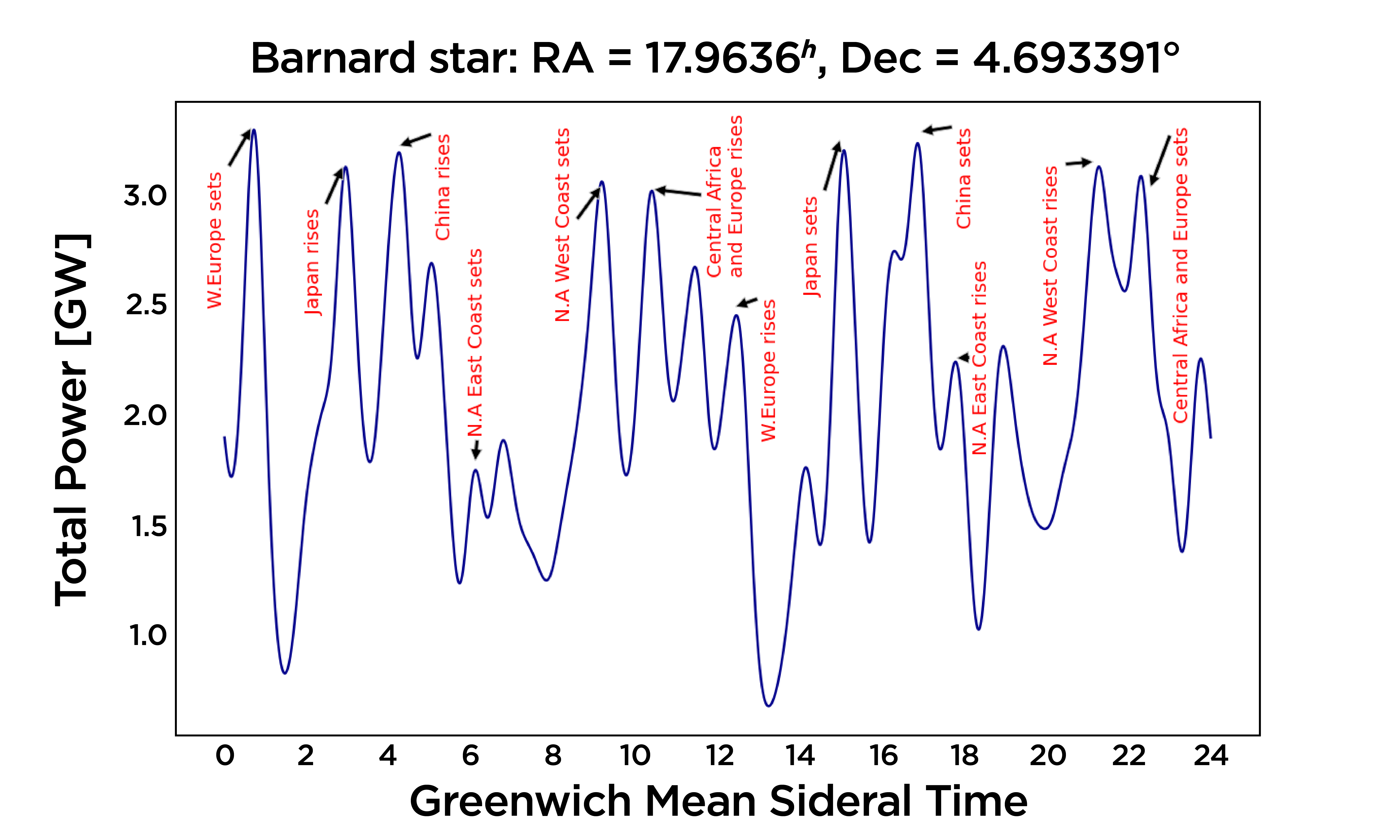}
\caption{Total Power of LTE mobile tower's leakage radiation plotted over a sidereal day in the direction of Barnard's star.}
\label{power_lte}
\end{figure*}

From figure \ref{power_lte} we can see that the contribution of these power levels comes mainly from Western Europe, in particular the regions of France, Belgium, the United Kingdom, the north of Spain, Germany, and Denmark. East Asia follows, with the main contribution coming from China, Japan, North and South Korea, Vietnam, and Australia. We also have significant leakage from the West of North America. The peaks represent the times at which Barnard's star rises and sets from these locations. From the star's frame of reference, the peaks occur when either Western Europe or East Asia first comes into view or finally disappears around the limb of the Earth (the edge of the Earth's disk) as seen from the star.

\begin{figure*}
\centering
\includegraphics[width=1\textwidth]{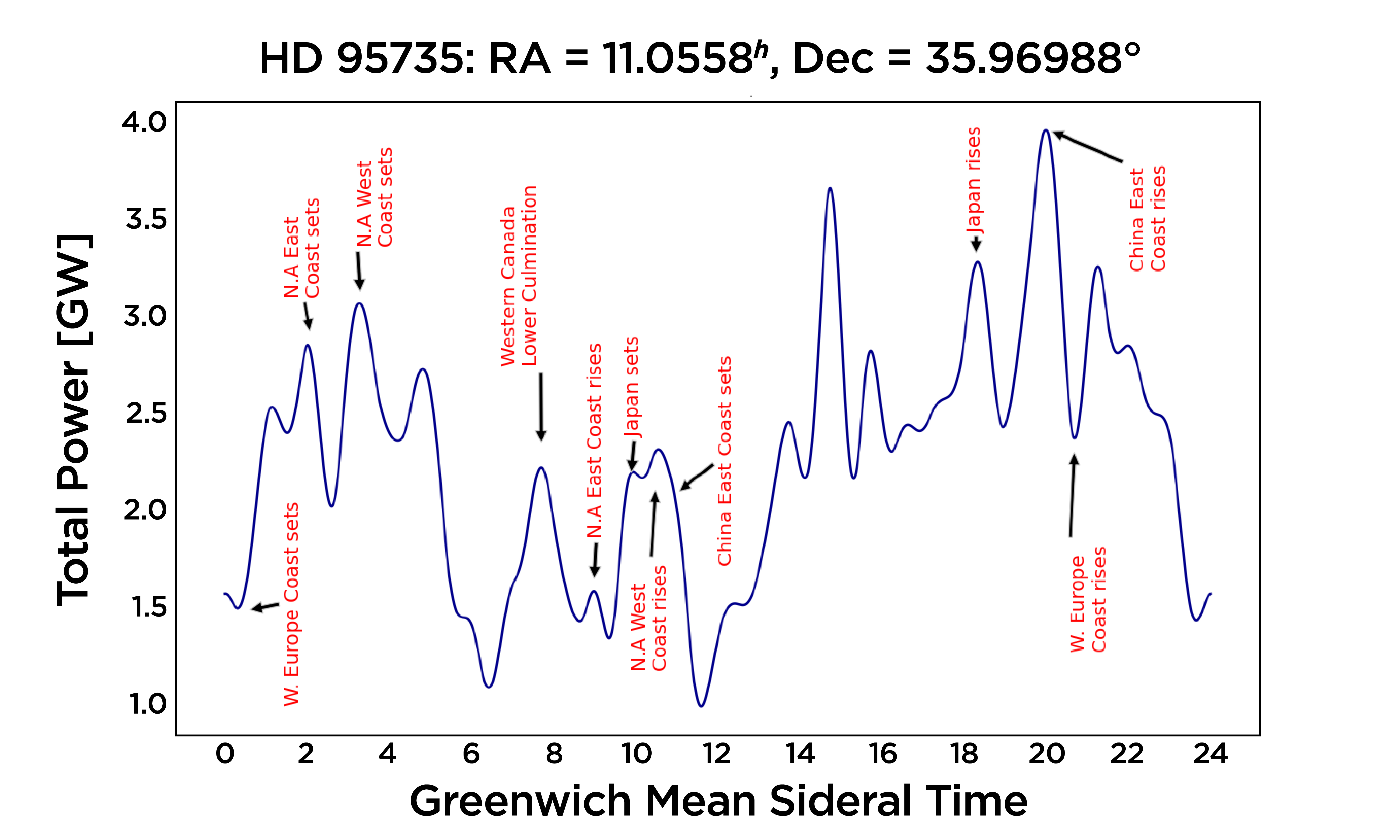}
\caption{Total Power of LTE mobile tower's leakage radiation plotted over a sidereal day in the direction of HD~95735 star.}
\label{HD95735}
\end{figure*}

The second hypothetical extraterrestrial observer that we chose was HD~95735, a red dwarf with Right Ascension: $11^{h} 03^{m} 20.1948^{s}$ and Declination: $+35^\circ 58' 11.5761"$ (J2000). This star is located $\sim 8.3$ light years away from the Earth \citep{gatewood1974astrometric}. Fig.\ref{HD95735}  represents the power structure of leakage caused by LTE mobile technology. From Fig.\ref{HD95735}, we can see that the contribution of these power levels comes mainly from the East Coast of China, followed by the West and East Coast of North America. The peak power leaking in the direction of HD 95735 is $\sim 4$~ GW. For the total power from GSM mobile tower emissions, we see the maximum power is $\sim 1.6$ ~GW, and for UMTS it is $\sim 3.3$~GW.

\begin{figure*}
\centering
\includegraphics[width=1\textwidth]{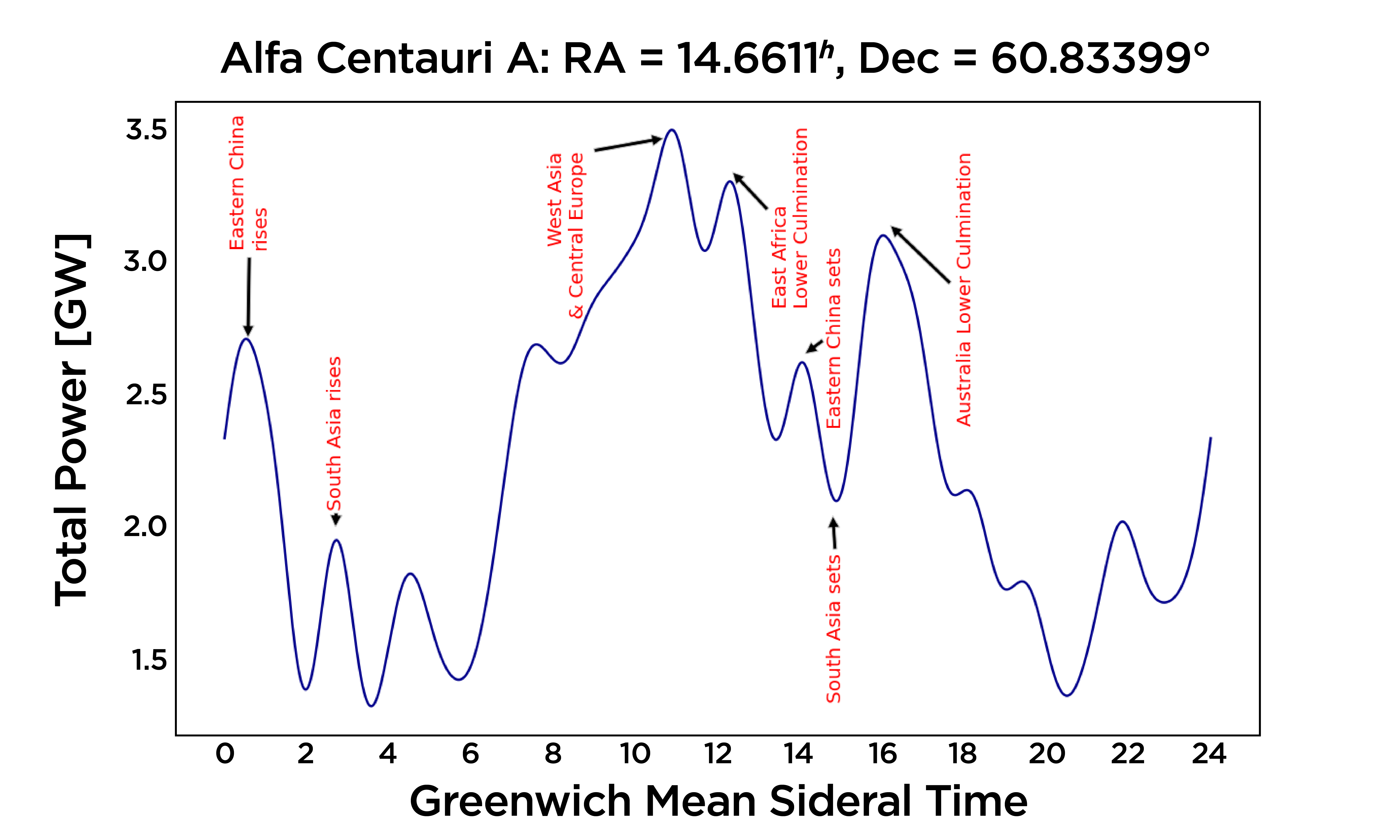}
\caption{Total Power of LTE mobile tower's leakage radiation plotted over a sidereal day in the direction of Alpha Centauri A.}
\label{alfa}
\end{figure*}

The third hypothetical observer was placed at Alpha Centauri A, with coordinates (RA = $14^{h} 39^{m} 36.4940^{s}$ and Dec = $-60^\circ 50' 02.3737"$ (J2000). This star is located at $\sim 4.2$ light years away from Earth \citep{zhao2017planet}
and we can see from figure \ref{alfa} that the contribution of the peak power levels comes mainly from the west of Asia and Central Europe, with East Africa and Australia also making significant contributions. Note that the main contribution from Australia occurs at lower culmination when this circumpolar star reaches its minimum altitude.  
The peak power leaking in the direction of Alpha Centauri A is approximately 3.5 GW. The total power from GSM and UTMS mobile towers is $\sim 1.4$~GW and $\sim 2.9$~GW respectively.

\section{Discussion} \label{section4}

Our findings are similar to the results presented earlier by \citet{sullivan1978eavesdropping}, demonstrating that the Earth's radio leakage remains periodic, including now the contribution made by mobile communication towers. This is not unexpected, since the Earth rotates and the physical distribution of towers across the surface of the planet is non-uniform in nature. The distributed location of mobile towers on the Earth, leads to a very complex variation in the planet's radio leakage signature, and from our results, we can see that it is also very dependent on the location of the hypothetical observer. In particular, the maximum leakage emission would be detected from northern stars (e.g. see HD~95735) - this is due to the fact that most mobile towers are located in the northern hemisphere. We also note that for low declination stars only a few transmitters on the Earth illuminate the extraterrestrial observer. For example, for an observer with  Dec = -89.2291 deg (e.g. HD 136509) \citep{2018yCat.1345....0G} and below, no transmitter illuminates the observer, nevertheless, the observer would still receive some amount of radiation. The same is true for stars with positive declination with values close to 90 degrees. 

Our study also draws attention to the fact, that the leakage signature of the Earth has evolved quite rapidly over relatively short time scales. The radio emission \citet{sullivan1978eavesdropping} estimated from TV transmitters and radar systems was dominated by a few thousand, very powerful transmitters located in specific areas of the planet. We deal with millions of mobile towers, operating at higher frequencies and employing much more modest powers. But perhaps the most interesting difference is that these towers are much more geographically distributed than the TV transmitters \citet{sullivan1978eavesdropping} considered. For the first time, our results (see Fig. \ref{alfa}) highlight the significant leakage contributions being made via the rise of developing countries in the continent of Africa, as well as by countries such as  Japan, Vietnam, China.

In this study, we have focused on leakage radiation from mobile towers. We have not included the leakage emission from mobile handsets themselves. The emission from handsets varies significantly, depending on whether they are actively transmitting or not. When a mobile phone is active the operator’s network controls the output power down to levels as low as 1~mW \citep{lonn2004output}. However, the sheer number of handsets active around the world at any given time, suggests this component should not be neglected. Since most mobiles are being used in densely populated areas, the bulk of handsets will be operating at fairly low power levels. We estimate the total background leakage emission from mobile handsets located around the Earth to be around an order of magnitude less powerful than the peak leakage produced by mobile towers presented here. Since mobile handsets radiate isotropically, they add a background component to the leakage that will be less variable than mobile towers since the latter are beamed towards the horizon. For the moment, we have not included mobile handsets in the analysis presented here. Our simulations therefore present a lower limit on the leakage radiation emanating from the Earth due to mobile communication systems. 

We note that by analysing the flux variation of our planet as a function of time, it should be possible for an extraterrestrial civilisation to generate a simple model of our planet that reproduces regions that are dominated by land, vegetation, and oceans/ice.  

\subsection{Detectability range}

In order to test whether these signals can be detected by an external observer, we first assume that they posses the same radio observing capabilities as we do.
The overall likelihood of detecting our signals from space depends on the frequency, transmitted power, bandwidth, the sensitivity of telescope, distance of the observer, the persistence of the leakage, etc. \citep{grimaldi2018bayesian}. 
For a given radio telescope, the minimum detectable flux density 
$S_{min}$ is given by:

\begin{equation}
    S_{min,wide}=SNR_{min}\frac{2k{T_{sys}}}{A_{eff}\sqrt{n_{pol}\Delta{t_{obs}}\Delta{\nu}}}
	\label{eq:quadratic}
\end{equation}

where $S/N_{min}$ is the signal-to-noise threshold value,$\Delta{t_{obs}}$ is the observing time, $\Delta{\nu}$ is the bandwidth, and $n_{pol}$ is the number of polarizations. The term 2k${T_{sys}}/A_{eff}$ is also known as SEFD. The SEFD is expressed in Jy (\(1 Jy = 10^{-26} W m^{-2} Hz^{-1})\). 

As a proxy to what our potential observer might have at their disposal, we will use in the first instance the Green Bank telescope (GBT). For the GBT at L-band, the SEFD is approximately 10~Jy \citep{enriquez2017breakthrough}.

The minimum EIRP that can be detected is then:

\begin{equation}
EIRP_{min} = 4\pi d^2 F_{min}
	\label{equation2}
\end{equation}

where $d$ is the distance to the transmitter, and $F_{min}$ is the minimum flux (in units of $W m^{-2}$) that can be detected by the observing system. The minimum detectable flux $F_{min} = S_{min} \delta\nu_{t}$, where $S_{min}$ is minimum detectable flux density and $\delta\nu_{t}$ is the bandwidth of our transmitted signal \citep{price2020breakthrough}. Assuming, $SNR_{min}$ = 7, $n_{pol}$ = 2, $\delta\nu_{t}$ = 20~MHz, $\Delta{t_{obs}}$ = 5 minutes, at the distance of 10 light-years, the $EIRP_{min}$ that can be detectable is $1.436\times10^{13} W$. 

For the Square Kilometre Array (SKA), the expected SEFD of SKA1-mid is $\sim 1.55$~Jy \citep{pellegrini2021mid}, and we can assume the full SKA (SKA Phase 2) would be a factor of 10 times better than that. These next generation radio telescopes would therefore be able to detect leakage radiation that is 1-2 orders of magnitude fainter than the GBT currently can. While these detection levels are still several orders of magnitude short of being able to detect our own peak leakage levels associated with current generation mobile systems, we might expect those leakage levels to increase in the future - for example, mobile towers for 5G communication systems are expected to operate at power-levels similar to 4G systems, these transmissions will occupy a much larger part of the radio spectrum in terms of the bandwidth they use. The flux density, measured by an extraterrestrial observer with a broad-band receiver system will therefore be significantly greater. Our results, also show that the regional distribution of mobile towers suggests that employing telescope integration times significantly greater than 5 minutes is also appropriate. Integration times of several hours are warranted from our plots presented in section 3, further increasing the sensitivity level by a factor of $\sim 5$. Taking all of this into account, the full SKA should be able to detect levels of $EIRP_{min} \sim 3\times10^{11} W$ - not so far removed from the power levels associated with our current 4G peak leakage signature $\sim 4\times10^{9}$~W.       

One aspect of detectability we have not yet considered is whether the leakage radiation from mobile towers can be distinguished against the powerful broad-band radio emission produced by the Sun. The Sun is a powerful source of radio waves over a wide range of different frequencies \citep{raulin2004analysis, raulin2005solar, tapping201310}. The solar radiation in the radio region can vary significantly when the Sun is active. During active periods, the solar flux reaches 1000 or even 10000 solar flux units (sfu - corresponding to $10^{-22} W m^{-2} Hz^{-1}$) at approximately 3 GHz. For an extraterrestrial observer, solar radio emission can be several orders of magnitude more powerful than the mobile tower emission \citep{muratore2022solar}. So a single dish radio telescope operated by a distant civilisation would have a tough time distinguishing between the two components of emission using broadband radiometer measurements. An interferometer, however, with baselines of several thousands of km would be able to resolve the Earth and Sun spatially, out to distances of $\sim 1$~kpc. Our hypothetical observer located at Barnard star, would spatially resolve the Sun and Earth with an L-band interferometer of baseline length 100 km. 
In this case an interferometer with sufficient sensitivity, would be able to detect a mobile leakage signal independently from the solar radio background which would also be resolved.

\subsection{Short-term evolution of mobile tower leakage }

It is expected that 5G mobile technology will account for over half of total mobile connections in developed regions of Asia and North America by 2025 \citep{statist}. Leakage power levels will increase but the services will also expand towards higher frequencies - in particular, 5G technology will operate in the following frequency bands: low-band (600, 800, and 900 MHz); mid-band (2.5, 3.5, and 3.7 – 4.2 GHz); and high-band (24, 26, 28, 37, 39, 42 and 47 GHz). Frequencies above 95 GHz are not yet licensed \citep{9115853}. With the anticipated rapid expansion of 5G technology, the amount of wireless devices is also expected to increase - resulting in a high density of infrastructure and broadband emissions. Higher frequency emissions have shorter ranges, thus a higher density of mobile towers will be needed, and this will increase the overall mobile tower leakage, as well as change its spectral profile. The rules adopted by the FCC allow a 5G base station operating in the millimeter range to emit an effective radiated power of up to 30,000 watts per 100 MHz of spectrum \footnote{shorturl.at/afKOP}. The nature of 5G leakage radiation will become clearer as it continues to be developed and deployed.

\section{Conclusions and Future work} \label{section5}

The main goal of the current study was to determine the power spectrum of mobile towers on Earth as observed by a hypothetical civilisation located at interstellar distances. Our findings show that the leakage radiation from mobile towers is variable in intensity and periodic in nature due to their non-uniform distribution on the Earth's surface and the rotation of our planet. We have simulated these effects in detail taking into account the position of mobile tower transmitters and the coordinates of the observer. Our results demonstrate that the maximum power generated by LTE mobile tower technology is of the order of 4~GW for HD 95735. By comparison, the UMTS mobile tower technology generates a power level of order 3.3~GW for HD~95735. The second most powerful emission leaks in the direction of Alpha Centauri A, is of order 3.5~GW, and corresponds also to LTE mobile technology. 

In terms of detectability, we conclude that any nearby civilization located within 10 light years of Earth and equipped with a receiving system comparable to the GBT would not detect the Earth's mobile tower leakage. Reciprocally we conclude that any existing extraterrestrial civilization leaking radio signals from mobile towers at the same power levels would not be detected by the GBT. Next generation telescopes such as the SKA can do better but are still some way off detecting these low levels of leakage emission. However, mobile systems are in their infancy, and the future development of this technology (e.g. 5G systems and beyond) suggests that this component of the Earth's leakage will continue to increase in power over time. If the leakage can be detected, an extraterrestrial observer would be able to discern various details of the nature of our planet and the distribution of technology on its surface. 

In the future, we plan to develop our model of mobile tower leakage  to include these more powerful and broader band 5G emissions. In addition, we would like to update and add other sources of human-made radio emission to our model, including both military and civilian radar systems, Deep Space Network (DSN) transmissions, communication satellites in both geostationary and low-earth-orbit, in particular the widely anticipated satellite constellations now planned e.g. Starlink and OneWeb \citep{mcdowell2020low, henri2020oneweb}. Last but not least, we would like to validate our results by comparing our simulations with real data - this would include radio telescopes observing  reflections from the Moon and satellites that monitor radio emissions from the Earth.

\section*{Acknowledgements}

Research reported is supported by a Newton Fund project, DARA (Development in Africa with Radio Astronomy), and awarded by the UK’s Science and Technology Facilities Council (STFC) - grant reference ST/R001103/1.

This research made use of Astropy, a community-developed core Python package for Astronomy \citep{astropy:2013, astropy:2018}.

\section*{Data Availability}

 The data underlying this article are available in \url{https://github.com/mirosaide/Qgis_selectbylocation.git} and \url{https://github.com/mirosaide/Mobile-Towers-Power-Calculation.git}, at https://dx.doi.org/[doi]. The datasets were derived from sources in the public domain: OpenCelliD, at \url{https://www.opencellid.org} and Natural Earth at \url{https://www.naturalearthdata.com/downloads/}.



\bibliographystyle{mnras}
\bibliography{bibliography} 








\bsp	
\label{lastpage}
\end{document}